\documentclass[11pt]{article}
\usepackage{amsmath,amsfonts,sint,epsfig,cite,graphics}
\usepackage{mathrsfs}
\usepackage{macros_static}
\usepackage{color}
\usepackage[english]{babel}
\usepackage{jheppub}

\usepackage{psfrag}
\usepackage{amsfonts}

\newcommand{\bdm}{\begin{displaymath}}
\newcommand{\edm}{\end{displaymath}}
\renewcommand{\be}{\begin{equation}}
\renewcommand{\ee}{\end{equation}}
\renewcommand{\bi}{\begin{itemize}}
\renewcommand{\ei}{\end{itemize}}

\begin{document}
\renewcommand{\vec}[1]{{\bf #1}}
\newcommand{\PH}{P_{\rm H}}
\newcommand{\PL}{P_{\rm L}}
\newcommand{\unity}{{I}}

\def\slash#1{\mkern-1.5mu\raise0.6pt\hbox{$\not$}\mkern1.2mu #1\mkern 0.7mu}
\begin{titlepage}

\begin{flushright}
\vskip .4cm
HIM-2011-03\\
MKPH-T-11-16\\
CERN-PH-TH/2011-306\\
\end{flushright}

\vskip 1.95cm
\begin{center}
{\Large\bf 
Towards a precise lattice determination of the leading hadronic 
contribution to $(g-2)_\mu$ 
}
\end{center}
\vskip 1.1cm
\begin{center}
{
Michele~Della~Morte$^{\scriptscriptstyle a}$, 
Benjamin J\"ager$^{\scriptscriptstyle a}$,
Andreas J\"uttner$^{\scriptscriptstyle b}$ and 
Hartmut Wittig$^{\scriptscriptstyle a}$

\vskip .5cm
}
{
\vskip 2.0ex
$^{\scriptstyle a}$
Institut~f{\"u}r~Kernphysik and Helmholtz Institut Mainz, \\
Johannes Gutenberg-Universit{\"a}t, D-55099 Mainz, 
Germany
\vskip 2.0ex
$^{\scriptstyle b}$
CERN, Physics Department, TH Unit, CH-1211 Geneva 23, Switzerland
}
\vskip .5cm
E-mail:\texttt{
morte@kph.uni-mainz.de, jaeger@kph.uni-mainz.de, 
Andreas.Juttner@cern.ch, wittig@kph.uni-mainz.de}
\end{center}
\vskip 1.cm
{\bf Abstract:}
We report on our computation of the leading hadronic contribution to the
anomalous magnetic moment of the muon using  two 
dynamical flavours of non-perturbatively O($a$)~improved 
Wilson fermions. The strange quark is introduced in the
quenched approximation.
Partially twisted boundary conditions are applied to improve the 
momentum resolution in the relevant integral.
Our results, obtained at three different values of the lattice spacing, allow 
for a preliminary study of discretization effects. We explore a wide range
of lattice volumes, namely 2~fm~$ \leq L \leq$~3~fm, with pion masses 
from 600 to 280 MeV and 
discuss  different chiral extrapolations to the physical point. We
observe a non-trivial dependence of $a_\mu^{\rm HLO}$ on 
$m_\pi$ especially for small pion masses.
The final result, $a_\mu^{\rm HLO}=618(64) \times 10^{-10}$, 
is obtained by considering only the quark connected contribution to the 
vacuum polarization.
We present a detailed analysis of systematic errors and discuss how they
can be reduced in future simulations.

\end{titlepage}

\newpage
\tableofcontents

\section{INTRODUCTION}

The magnetic moment of a charged lepton is extracted from the vertex function 
describing the interaction between the lepton and a photon in the limit
of vanishing photon momentum. The corresponding anomalous magnetic moment $a_l$ 
is then defined as half the difference between the gyromagnetic factor $g$ and 
its classical value of 2, i.~e. $a_l=(g_l-2)/2$.
In the case of the electron the quantity is dominated by QED contributions.
The one-loop result was obtained by Schwinger more than sixty 
years ago~\cite{Schwinger:1948iu}, and since then $a_e$ has reached an
accuracy better than one part per billion on both the theoretical and 
the experimental sides, which yield results in beautiful agreement (see section 3 of~\cite{Jegerlehner:2009ry}, 
and references therein).

The anomalous magnetic moment mediates
helicity flip transitions~\cite{Cowland:1958}, which implies that quantum corrections due
to heavier particles of mass $M$, in the Standard Model or beyond, are proportional
to $m_l^2/M^2$. For this reason the muon anomalous magnetic moment $a_\mu$ is
regarded as a sensitive probe for effects of nearby New Physics. 
However, by the same argument, given that $m_\mu \leq m_\pi$, the hadronic
contributions to $a_\mu$ are large and  notoriously difficult to quantify.
While the experimental and theoretical estimates have each reached similar 
levels of precision of 0.5 ppm, a tension by 2 or 3 standard deviations between theory 
and experiment  persists~\cite{Davier:2010nc,Jegerlehner:2011ti}.
Before invoking New Physics as the reason for this tension  
the theoretical result and, in particular, all contributions due to hadronic effects, 
must be corroborated.
The uncertainty is dominated  mainly by  
the hadronic leading order ($a_\mu^{\rm HLO}$) and secondly by the hadronic light-by-light contributions.
Currently $a_\mu^{\rm HLO}$ is estimated via a phenomenological approach 
based on the evaluation of a dispersion integral. In the low-energy regime the spectral 
function in the integrand must be determined experimentally, either from the cross section
$e^+e^- \to \; hadrons$ or from the rate of hadronic $\tau$-decays.
Both methods suffer from different systematics~\cite{Jegerlehner:2009ry,Jegerlehner:2011ti}
and yield results in slight tension among each other. None of them however reduces the discrepancy
between theory and experiment on $a_\mu$ below the two standard deviation level.
A purely theoretical estimate of  $a_\mu^{\rm HLO}$ from a first-principles approach is clearly desirable,
and the present work represents our first step in that direction by using the lattice regularization
of QCD.

The hadronic vacuum polarization contribution to $a_\mu$
 has received considerable attention from the lattice community in recent years.
Initial studies have been performed in the quenched approximation~\cite{Blum:2002ii,Gockeler:2003cw} and
in the theory with two~\cite{Feng:2011zk} and three dynamical flavours~\cite{Aubin:2006xv,Boyle:2011hu}.
Compared to the determination of ``standard'' quantities such as hadron masses, quark masses and decay constants
(see~\cite{Colangelo:2010et} for a review), lattice
calculations of $a_\mu^{\rm HLO}$ are extremely challenging. The relevant lattice quantity 
(the hadronic vacuum polarization discussed in the next section) 
receives contributions from quark disconnected
diagrams, which are intrinsically  noisy and difficult to estimate with good statistical accuracy
at a reasonable numerical cost. Moreover, the dependence of the hadronic vacuum polarization on the 
momentum transfer must be accurately traced down to
momenta of order $m_\mu^2$ and beyond. This value is well below the lowest 
Fourier momentum $(2\pi/L)^2$ which can be reached in current lattice QCD simulations.
As a consequence finite size effects may conceivably be large on results
obtained within the ``standard'' approach.
In addition, the vacuum polarization receives sizeable contributions from the 
low-lying vector resonances.
Those should be properly accounted for in simulations
at sufficiently light quark masses 
for the lowest vector meson to be a resonance 
and including the full dynamics of the strange and charm quarks. Finally, as the
contribution to $a_\mu^{\rm HLO}$ from isospin breaking effects may be of the
same size of its present uncertainty, those will have to be included
in lattice computations (possibly along the lines discussed in~\cite{Blum:2007cy,Blum:2010ym,deDivitiis:2011eh})
for them to have a crucial impact on $(g-2)_\mu$ phenomenology.

In~\cite{Juttner:2009yb,DellaMorte:2010aq} we have shown how (partially) twisted boundary 
conditions~\cite{deDivitiis:2004kq,Sachrajda:2004mi,Bedaque:2004ax} can be used
to improve the momentum resolution in the connected part of the hadronic vacuum
polarization, and we have obtained an estimate of the
disconnected contribution in Chiral Perturbation Theory, thus addressing the first 
two systematic effects discussed above. Here we report
on numerical results obtained in that setup and show how the 
fits to the momentum dependence of the vacuum polarization function get 
substantially stabilized in this way, thereby improving the accuracy of our estimates.
Partial results have already appeared in~\cite{DellaMorte:2010sw,DellaMorte:2011ge}. In addition
we present a thorough investigation of all sources of systematic errors, including the modelling
of the $q^2$-dependence of the vacuum polarization, chiral extrapolations, lattice artifacts and 
finite volume effects.

\section{DEFINITIONS AND LATTICE SETUP}

The Euclidean hadronic vacuum polarization (VP) tensor is defined as
\begin{equation}\label{eqn:pol_tensor}
\Pi_{\mu\nu}^{(N_{\rm f})}(q)=
\int d^4xe^{iqx}\langle
	J_\mu^{(N_{\rm f})}(x)J_\nu^{(N_{\rm f})}(0)
	\rangle 
\,,
\end{equation}
where $J_\mu^{(N_{\rm f})}(x)=\sum\limits_{f=1}^{N_{\rm f}}Q_f\psibar_f(x)\gamma_\mu \psi_f(x)$. For
$N_{\rm f}=2$ the quark fields $\psi_f(x)$ are taken from the set $(\psi_u,\psi_d)$ with $Q_f=(2/3,-1/3)$ denoting the 
electric charges in units of the elementary one. For $N_{\rm f}=2+1$,
	$f=(u,d,s)$ and $Q_f=(2/3,-1/3,-1/3)$.
Euclidean invariance and current conservation imply
\begin{equation}
\Pi_{\mu\nu}^{(N_{\rm f})}(q)=(g_{\mu\nu}q^2-q_\mu q_\nu)\Pi^{(N_{\rm f})}(q^2)\,.
\end{equation}
For space-like momenta, the relation between $\Pi_{\mu\nu}^{(N_{\rm f})}(q^2)$ and 
the lowest order hadronic contribution $a_\mu^{\rm HLO}$ 
to the anomalous magnetic moment of the 
muon has been derived in~\cite{deRafael:1993za,Blum:2002ii,Gockeler:2003cw} and reads
(suppressing the index $N_{\rm f}$)
\begin{equation}
a_\mu^{\rm HLO}= \left( {{\alpha}\over{\pi}} \right)^2 \int_0^\infty dq^2 \, f(q^2)
\hat{\Pi}(q^2) 
\;,
\label{eq:amu}
\end{equation}
where
\be
f(q^2)={{m_\mu^2q^2Z^3(1-q^2Z)}\over{1+m_\mu^2q^2Z^2}}\;, \quad \quad
Z=-{{q^2-\sqrt{q^4+4m_\mu^2q^2 }}\over{2m_\mu^2q^2}}\,,
\label{eq:f}
\ee
and $\hat{\Pi}(q^2)\equiv 4\pi^2\left[\Pi(q^2)-\Pi(0)\right]$.
\subsection{Lattice regularization}
We perform our computation on a subset of the
 gauge configurations generated within the CLS initiative~\cite{CLS} 
for two flavours of non-perturbatively O$(a)$ improved Wilson
fermions~\cite{Jansen:1998mx} and using the standard Wilson plaquette gauge action.
The simulation parameters are collected in Table~1, including the values
of the twist angle $\theta$, whose r\^ole will be explained below.
\begin{table}[htb] 
\label{tab:par}
\begin{center}
\begin{tabular}{cccccccc}
    	\hline\hline
 	 &$T\cdot L^3$ & $\beta$ & $m_\pi$ $[\mathrm{MeV}]$  & $a$ $[\mathrm{fm}]$ &$\theta$ & $\kappa_s$ & \# meas.\\
    	\hline
    	A3 & $64 \cdot 32^3 $ & 5.20 & 471 & 0.079 & 0.8; 1.8; 2.6 & - & 532\\
	A4 & $64 \cdot 32^3 $ & 5.20 & 362 & 0.079 & 0.8; 1.8; 2.6 & - & 800\\
	A5 &$64 \cdot 32^3 $ & 5.20 & 317 & 0.079 & 0.8; 1.8; 2.6 & - & 432\\
    	\hline
    	E4 & $64 \cdot 32^3 $ & 5.30 & 601 & 0.063 & 0.8; 1.8; 2.6 & 0.13605 & 648 \\
	E5 & $64 \cdot 32^3 $ & 5.30 & 447 & 0.063 & 0.8; 1.8; 2.6 & 0.13574 & 672\\
	F6 & $96 \cdot 48^3 $ & 5.30 & 324 & 0.063 & 0.4; 1.9; 2.3 & 0.13575 & 804\\
    	F7 & $96 \cdot 48^3 $ & 5.30 & 277 & 0.063 & 0.4; 1.9; 2.3 & 0.13570 & 820\\
    	\hline
    	N4 & $96 \cdot 48^3 $ & 5.50 & 541 & 0.050 & 0.8; 1.9; 2.6 & 0.13639 & 532\\
    	N5 & $96 \cdot 48^3 $ & 5.50 & 431 & 0.050 & 0.8; 1.9; 2.6 & 0.13629 & 644\\
        \hline\hline	
\end{tabular}	
\caption{\small{Summary of simulation parameters. Measurements are performed on configurations
separated by 8 units of molecular dynamics time at least, where one molecular dynamic unit
corresponds to an evolution by $\tau=1$ in Hybrid Monte Carlo algorithms. The lattice spacings and 
the values of the hopping parameter $\kappa_s$ corresponding to the strange quark mass
are taken from~\cite{Capitani:2011fg} and have been determined using the definitions and some of the results 
in~\cite{DelDEbbio:2006cn,Capitani:2009tg,Brandt:2010ed}. The values of the lattice scale 
must be regarded as preliminary.
The masses of the lightest pseudoscalar
state $m_\pi$ are taken from~\cite{masses}.
The twist angle $\theta$ is applied
in the spatial $x$-direction only.}}
\end{center}
\end{table}

Following~\cite{Gockeler:2003cw}, we have implemented the one-point-split conserved vector current
\be
V^f_\mu(x)={{1}\over{2}}\left( \psibar_f(x+a\hat{\mu})(1+\gamma_\mu) U_\mu^\dagger(x) \,\psi_f(x) \;-\;
\psibar_f(x)(1-\gamma_\mu)U_\mu(x)\,\psi_f(x+a\hat{\mu})\right) \;,
\ee
where $U_\mu \in SU(3)$ represents the gauge link in the positive $\mu$ direction, and $f$ is  again
a flavour index.
The lattice version of the vacuum polarization tensor then reads
\be
\Pi^{(N_{\rm f})}_{\mu\nu}(x)=a^6{\Big{\langle}} \sum_{f=1}^{N_{\rm f}} (Q_fV^f_\mu(x)) 
 \sum_{f'=1}^{N_{\rm f}} (Q_{f'}V^{f'}_\nu(0)) {\Big{\rangle}} \;.
\ee
By Fourier transforming the expression above into momentum space one gets
\be
\Pi^{(N_{\rm f})}_{\mu\nu}(\hat{q})=\sum_x e^{iq(x+a\hat{\mu}/2 - a\hat{\nu}/2)} \Pi^{(N_{\rm f})}_{\mu\nu}(x)\;,
\label{eq:Foutra}
\ee
where $q_\mu=2 \pi n_\mu/L$ and $\hat{q}_\mu=\frac{2}{a}\sin\left(\frac{aq_\mu}{2}\right)$ with
 $n_\mu \in 0,1, \dots L/a-1$. We restrict our attention to the case $\mu \neq \nu$ to avoid
mixings of the composite field $V_\mu(x)V_\nu(0)$ with lower dimensional ones in the limit $x \to 0$.
It then follows that $\Pi^{(N_{\rm f})}_{\mu\nu}(\hat{q})$ fulfils the Ward identities
\be
\hat{q}_\mu \Pi^{(N_{\rm f})}_{\mu\nu}(\hat{q}) = \Pi^{(N_{\rm f})}_{\mu\nu}(\hat{q})\, \hat{q}_\nu =0\;.
\ee
Consequently on the lattice we can relate the scalar vacuum polarization to the tensor one,
by mimicking the continuum relation
\be
\Pi^{(N_{\rm f})}_{\mu\nu}(\hat{q}) = (\hat{q}^2 \delta_{\mu\nu}-\hat{q}_\mu \hat{q}_\nu)\Pi^{(N_{\rm f})}(\hat{q}^2)\;.
\ee
The scalar vacuum polarization extracted in this way  approaches its continuum counterpart
with a rate proportional to $a$, where the O$(a)$  effects
appear due to off-shell contributions in $\Pi^{(N_{\rm f})}_{\mu\nu}(\hat{q})$.

After performing the Wick contractions in $\Pi_{\mu\nu}^{(N_{\rm f})}(x)$ one realizes that
connected as well as disconnected quark diagrams contribute. We neglect here the disconnected terms,
as done in most of the existing lattice computations in the literature. In~\cite{Feng:2011zk} such contributions
were included for almost half of the ensembles used and found to be negligible within the quoted errors.
Following the arguments in~\cite{DellaMorte:2010aq}, the disconnected diagrams are expected to decrease the value of 
$\hat{\Pi}_{\mu\nu}^{(N_{\rm f})}(\hat{q})$ by about 10\%, which 
demonstrates the limited accuracy of lattice calculations without their proper inclusion.
\subsection{Twisting the connected part of the vacuum polarization}

By modifying the spatial boundary conditions on the quark fields entering the vector
current it is possible to improve the momentum resolution in $\Pi^{(N_{\rm f})}(\hat{q}^2)$
and to access momenta different from the integer multiples of $2\pi/L$.
Namely, imposing the condition
\be
\psi(x+L\hat{k})=e^{i\theta_k}\psi(x) \; 
\ee
is equivalent to boosting the momenta in the quark propagator by $\theta_k/L$.
This technique can be used to modify and refine the lattice dispersion relation
of, for example, charged pseudoscalar mesons, or of any momentum dependent quantity,
as long as there are no strong final state interactions~\cite{Sachrajda:2004mi}. 
This remains true, up to exponentially suppressed finite size effects, 
 for partially twisted boundary conditions~\cite{deDivitiis:2004kq,Sachrajda:2004mi,Bedaque:2004ax},
where only (some of) the valence quarks satisfy twisted boundary conditions, while the sea quarks
fulfil periodic boundary conditions. Note, however, that the effect of partial twisting obviously vanishes
whenever it is applied to flavour-singlet quantities, as the vector current discussed above.

We have shown in~\cite{DellaMorte:2010aq} that, by introducing a sufficiently large number
of valence quarks degenerate with the $u,\,d \dots$ flavours, the quark connected and disconnected
parts of the correlator $\Pi_{\mu\nu}^{(N_{\rm f})}(x)$ can be rewritten as independent correlation
functions in Partially Quenched QCD. In such unphysical theories each quark diagram has 
an unambiguous field-theoretic expression and well-defined continuum and infinite volume limits.
The formulation naturally
turns the connected contribution into a correlator of flavour-off-diagonal vector currents and thus 
twisting can be applied to induce arbitrary momentum. This amounts to a simple modification of
Eq.~\ref{eq:Foutra} (restricted to its connected part) in which $q_k \to q_k-\theta_k/L$ and
$\hat{q}$ changes accordingly. 

In practice we have applied twisting in the $x$-direction only 
and remained with periodic boundary conditions in the other directions.
The twist angles in Table~1 have been chosen in order to achieve a fixed and large density
of $q^2$ values between consecutive Fourier modes and in order to reach a lowest non-zero
$q^2$ around $m_\mu^2$. For the Fourier modes we have considered all the integers values
between $(0,0,0,0)$ and $(2,2,2,2)$ in units of $2\pi/L$.

\section{RESULTS}
In order to compute $a_\mu^{\rm HLO}$ we must obtain
a continuous description of $\Pi^{(N_{\rm f})}(\hat{q}^2)$ to determine 
$\Pi^{(N_{\rm f})}(0)$ and 
$\hat{\Pi}^{(N_{\rm f})}(\hat{q}^2)$ entering the 
integral in Eq.~\ref{eq:amu}. At each value of the lattice spacing $a$
we introduce a maximum momentum $\hat{q}^2_{\max}$ and adopt several 
 different  fit {\it ans\"atze} for the low $\hat{q}^2$ region below
$\hat{q}^2_{\max}$. We additionally impose a matching condition 
with perturbation theory at  
$\hat{q}^2=\hat{q}^2_{\max}$ and hence use the perturbative expression
to describe the high $\hat{q}^2$ region.

The results for  $a_\mu^{\rm HLO}$ at different values of the mass of the
lowest pseudoscalar state $m_{\pi}$ are then extrapolated to the physical point
 using a functional form inspired by Chiral Perturbation Theory.
\subsection{Fitting procedure}
\begin{figure}[htb]
\begin{center}
\includegraphics[width=10cm]{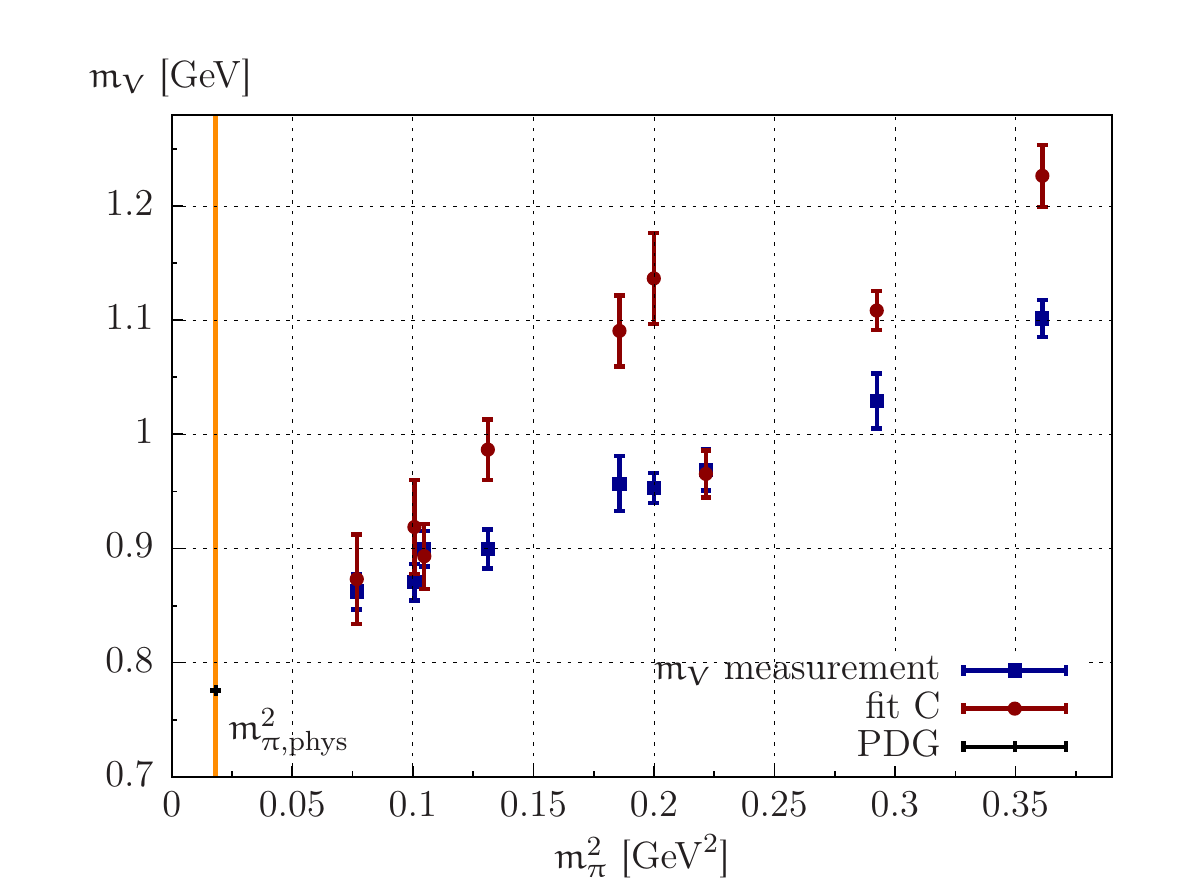}
\caption{\small{Comparison between the smaller mass parameter from fit {\it ansatz} C and the vector mass measured
from the vector two-point function as a function of $m_\pi^2$.}}
\label{mv}
\end{center}
\end{figure}

In order to check for the stability of our results against variations in the fitting procedure
we use three different  {\it ans\"atze} to describe $\Pi^{(N_{\rm f})}(\hat{q}^2)$
in the region $0 \leq \hat{q}^2 \leq \hat{q}^2_{\rm max}$:
\begin{itemize}
\item[A)] a model-independent $(2,3)$ Pad\'e {\it ansatz} defined by a degree 2 over a degree 3 polynomial in $\hat{q}^2$ ($n_{\rm coeff}=6$), i.e.
\be
\Pi^{(N_{\rm f})}(\hat{q}^2)={{a(b^2+\hat{q}^2)(c^2+\hat{q}^2)}\over{(d^2+\hat{q}^2)
(e^2+\hat{q}^2)(f^2+\hat{q}^2)}}\;,
\label{eq:fitA}
\ee
\item[B)] a vector-dominance model with a single pole $(n_{\rm coeff}=3)$, i.e.
\be
\Pi^{(N_{\rm f})}(\hat{q}^2)=a + {{b}\over{c^2+\hat{q}^2}}\;,
\label{eq:fitB}
\ee
\item[C)] a vector dominance-model with two poles\footnote{Up to a re-definition of the fitting parameters the
{\it ans\"atze} B and C are equivalent to Pad\'e approximations of degree $(1,1)$ and $(2,2)$ respectively.}. 
This form ($n_{\rm coeff}=5$) has been suggested and used in~\cite{Boyle:2011hu} and reads
\be
\Pi^{(N_{\rm f})}(\hat{q}^2)=a + {{b}\over{c^2+\hat{q}^2}}+ {{d}\over{e^2+\hat{q}^2}}\;,
\label{eq:fitC}
\ee
subject to the additional constraint $c^2<e^2$.
\end{itemize}
We perform correlated fits. For these to be reliable and to produce a 
meaningful correlated $\chi^2$ large statistics is needed.
In~\cite{Michael:1993yj} a toy model was used to provide the thumb-rule $N \,\grtsim \,D^2$, where $N$ is the number of 
measurements and $D$ the number of degrees of freedom in the fit.
We therefore randomly select about 25 values of $\hat{q}^2$ and indeed observe stable 
correlated fits as well as the absence of  numerical problems when inverting the average covariance matrix.
The errors on the fit parameters are estimated and propagated to 
$a_\mu^{\rm HLO}$ by performing the fits and the  integral 
for each bootstrap sample.
For the fit {\it ansatz} C we compare in Fig.~\ref{mv} the value of the smaller mass
parameter obtained from the fit with the mass measured from the exponential decay of
the vector two-point function. We call the latter {\it naive} vector mass, as it
assumes the lightest vector meson to be stable  at all values of $m_\pi$ considered.
While this may not be the case for our most chiral points, for the sake of this
qualitative comparison such a definition seems nonetheless adequate. A separate study will have to be
devoted to the measurements of the vector resonance parameters following~\cite{Luscher:1985dn,Luscher:1986pf}. 
The figure shows overall reasonable agreement, with discrepancies which
stretch to 15\% at most.

As a constraint on the high-$\hat{q}^2$ region
we always include perturbation theory in the $\overline{\rm MS}$
scheme at the scale 2~GeV, using the  leading-order expression
in $\alpha_{\rm s}$ for $\Pi^{(N_{\rm f})}(q^2)$
from~\cite{Chetyrkin:1996cf}. Although the NLO in $\alpha_{\rm s}$ is available
from the same reference, for our purposes it is enough to include the leading-order,
where in addition no internal fermionic loops appear and therefore it is rather
simple to apply the formulae to the case of two light dynamical quarks and a quenched
strange. In addition, as it will become clear in the following, the perturbative contribution
 to $a_\mu^{\rm HLO}$ estimated in this way is extremely small.
To evaluate the perturbative
formula we use the non-perturbative 
$\Lambda_{\overline{\rm MS}}$ parameter for $N_{\rm f}=2$ from~\cite{DellaMorte:2004bc} and
the non-perturbative renormalization factors 
in~\cite{DellaMorte:2005kg,DellaMorte:2008xb,Fritzsch:2010aw} 
to relate the lattice quark masses to their values in the $\overline{\rm MS}$ 
scheme at 2~GeV.
As discussed in~\cite{Gockeler:2003cw} the function $\Pi(\hat{q}^2)$ computed on the lattice and in the
continuum dimensional regularization can differ by an un-physical and non-universal integration constant.
We fix the value of this scheme dependent constant
by matching the perturbative expression for $\Pi(\hat{q}^2)$ to our non-perturbative data
at $\hat{q}^2=\hat{q}^2_{\rm max}$.
Moreover the perturbative prediction diverges as $q^2\to 0$ and
therefore even after this step  perturbation theory can be expected to 
provide a good description of our data only at sufficiently large values of
$\hat{q}^2$. 
We require the resulting function not only to be smooth but also to have a 
smooth first derivative on the whole real axis, and in particular at $\hat{q}^2=\hat{q}^2_{\rm max}$.
Hence, the matching produces one non-trivial relation among the parameters in  the fit {\it ans\"atze} above,
reducing their number $n_{\rm par}$ from $n_{\rm par} = n_{\rm coeff}$ to $n_{\rm coeff}-1$.

\begin{figure}[htb]
\begin{minipage}{0.49\linewidth}
	\centering
	\includegraphics[width=\linewidth]{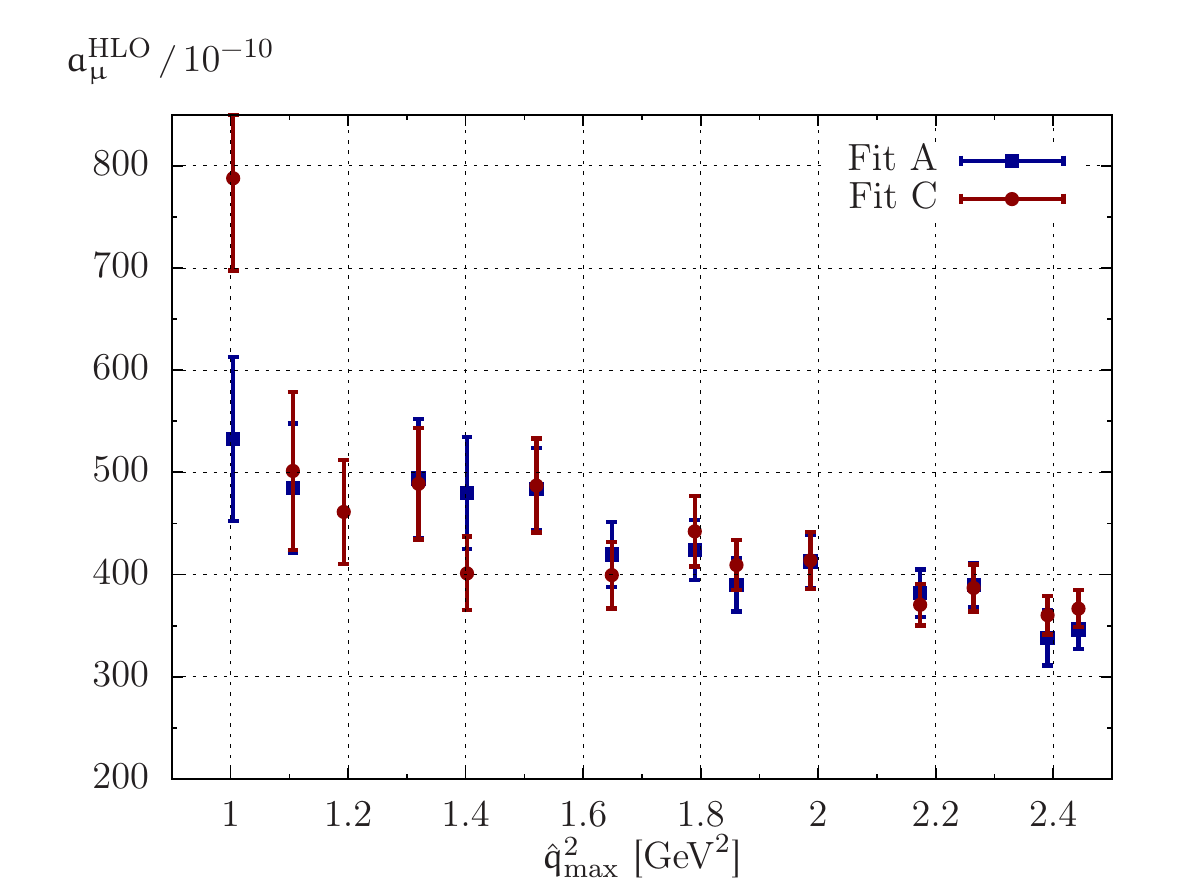}
\end{minipage}
\hfill
\begin{minipage}{0.49\linewidth}
	\centering
	\includegraphics[width=\linewidth]{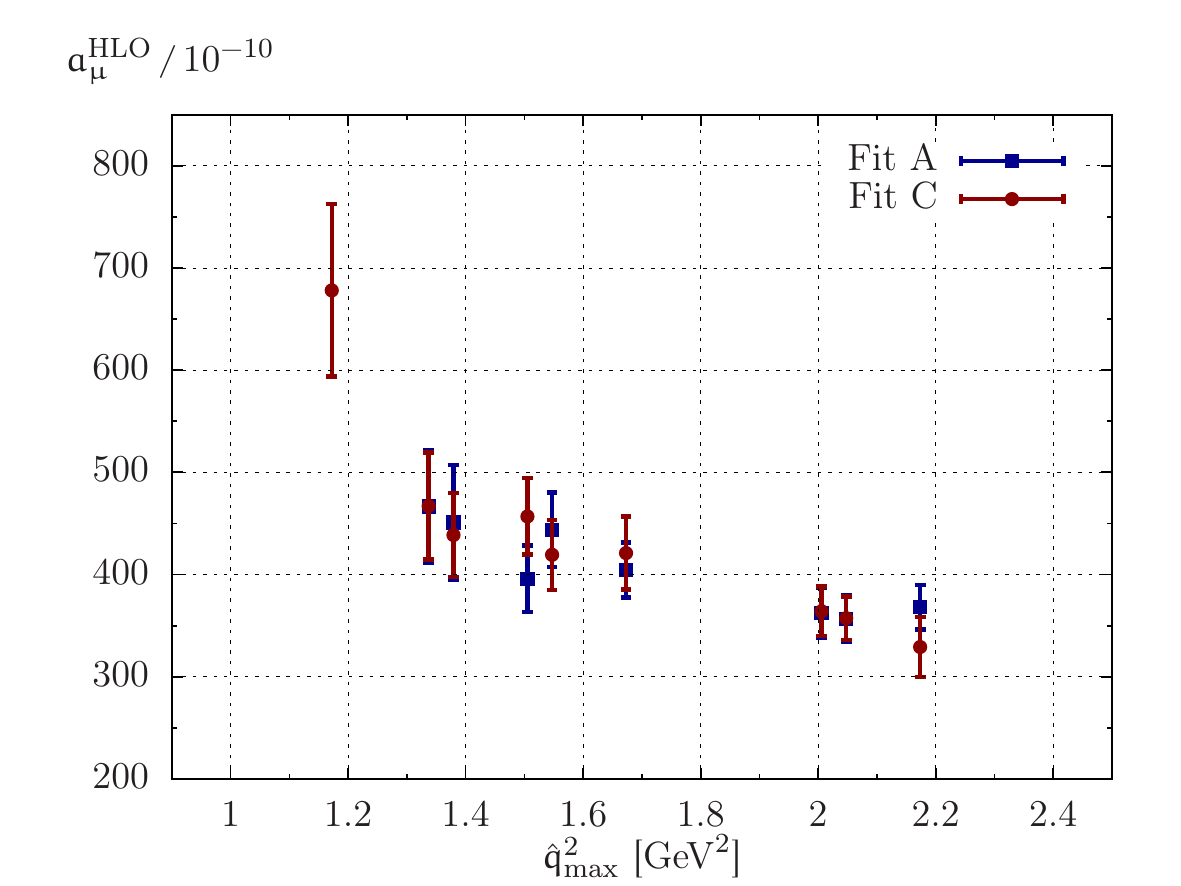}
\end{minipage}
\caption{\small{{\bf Left:}
Fit results from the F6 ensemble using the {\it ans\"atze} A and C in the text, plotted as a function
of $\hat{q}^2_{\rm max}$. {\bf Right:} Equivalent plot obtained by using the Fourier momenta only.}}
\label{fits}
\end{figure}

The fit {\it ansatz} B always produces reduced $\chi^2$  larger than 5 and we therefore
discard it. We monitor the stability of the fit results for $a_\mu^{\rm HLO}$  in the way 
shown in the left panel of Fig.~\ref{fits} where, as an example,  the values obtained from fits of type A and C 
to the measurements performed on the F6 ensemble are plotted. 
A cut $\chi^2/{\rm dof} \leq 2.5 $ is applied in producing the figure.
The two functions give compatible results for $\hat{q}^2_{\rm max} \, \grtsim \,1.5 \,{\rm GeV}^2$ 
and in the end we opt for the results  from {\it ansatz} A  
on all the ensembles, as this yields in most cases the smallest $\chi^2$ values and 
the smallest errors on $a_\mu^{\rm HLO}$. The values of $\hat{q}^2_{\rm max}$ are chosen 
differently for different ensembles as the $\hat{q}^2$-ranges explored
depend on the lattice spacing and on $L/a$.
Specifically the maximum momentum lies in the range 
$2.4 \,{\rm GeV}^2\,\ltsim \;\hat{q}^2_{\rm max} \;\ltsim \,5.5 \,{\rm GeV}^2$.
In order to quantify the improvement brought by the use of partially
twisted boundary conditions we repeat in Fig.~\ref{fits}, right panel, the comparison between fit {\it ans\"atze} A and C
on the F6 ensemble but restricting the data to the Fourier modes only.
Twisting clearly helps in stabilizing the fits and the plateau values are reached "earlier"
in $\hat{q}^2_{\rm max}$. This increases our confidence in the fitting adopted procedure. 
Statistical errors are also slightly smaller in the left panel of Fig.~\ref{fits} with respect to those in the right one.

We summarize our results on $a_\mu^{\rm HLO}$ for  the  two-flavour theory 
and the case with an additional quenched strange quark in Tables~2~and~3
respectively.
\begin{table}[htb]
\label{tab:nf2}
\begin{center} 
\begin{tabular}{cclclc}
        \hline\hline
         & & \multicolumn{2}{c}{fit A} & \multicolumn{2}{c}{fit C}\\
	 &$\hat{q}_\mathrm{max}^2 \,[\mathrm{GeV}^2]$ & \multicolumn{1}{c}{$a_\mu^\mathrm{HLO}$} & $\chi^2 / \mathrm{dof}$  & \multicolumn{1}{c}{$a_\mu^\mathrm{HLO}$} & $\chi^2 / \mathrm{dof}$\\
    	\hline
	A3 & 3.5 & 274.2(9.8) & 0.96 & 282.4(9.1) & 0.97\\
	A4 & 3.5 & 304.2(13.5) & 1.25 & 305.0(12.8) & 1.12\\
	A5 & 3.5 & 395.8(40.2) & 0.84 & 384.3(30.1) & 0.95\\
	\hline
	E4 & 5.5 & 197.0(5.8) & 1.02 & 195.4(6.0) & 0.88\\
    	E5 & 5.5 & 248.5(10.8) & 0.83 & 255.6(13.6) & 0.92\\
	F6 & 2.4 & 346.3(19.0) & 1.05 & 366.7(18.3) & 1.19\\
	F7 & 2.4 & 406.8(37.6) & 1.15 & 382.8(24.5) & 1.22\\
	\hline
	N4 & 3.9 & 253.4(5.2) & 0.46 & 252.0(5.0) & 0.65\\
	N5 & 3.9 & 273.3(9.4) & 0.53 & 276.2(11.2) & 0.57\\
\hline\hline
\end{tabular}	
\caption{\small{Results for $a_\mu^{\rm HLO}$ in the $N_{\rm f}=2$ theory.}}
\end{center}
\end{table}
\begin{table}[htb]
\label{tab:nf2q} 
\begin{center}
\begin{tabular}{cclclc}
        \hline\hline  
         & & \multicolumn{2}{c}{fit A} & \multicolumn{2}{c}{fit C}\\
	 &$\hat{q}_\mathrm{max}^2 \,[\mathrm{GeV}^2]$ & \multicolumn{1}{c}{$a_\mu^\mathrm{HLO}$} & $\chi^2 / \mathrm{dof}$  & \multicolumn{1}{c}{$a_\mu^\mathrm{HLO}$} & $\chi^2 / \mathrm{dof}$\\
    	\hline
	E4 & 5.5 & 236.8(6.6) & 1.10 & 233.7(4.5) & 0.89\\
    	E5 & 5.5 & 294.8(15.4) & 0.89 & 291.3(12.3) & 0.72\\
	F6 & 2.4 & 404.5(19.7) & 1.29 & 403.1(20.5) & 1.39\\
	F7 & 2.4 & 457.5(28.5) & 1.13 & 452.1(28.1) & 1.31\\
	\hline
	N4 & 3.9 & 303.2(6.3) & 0.55 & 300.4(5.7) & 0.73\\
	N5 & 3.9 & 323.5(9.3) & 0.59 & 330.3(10.8) & 0.57\\
        \hline\hline
\end{tabular}	
\caption{\small{Results for $a_\mu^{\rm HLO}$ in the theory with two dynamical light quarks and a quenched strange quark.}}
\end{center}
\end{table}

\begin{figure}[htb]
\begin{minipage}{0.68\linewidth}
	\centering
	\includegraphics[width=\linewidth]{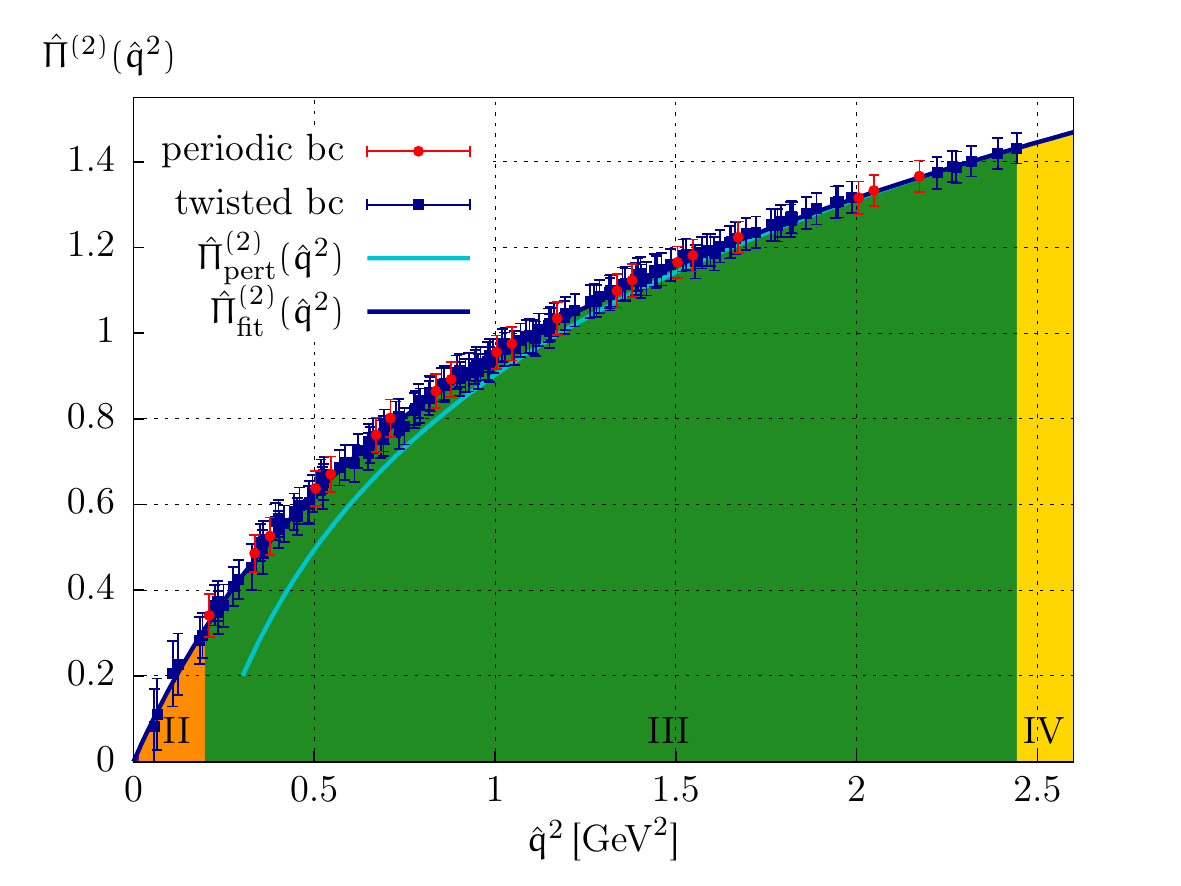}
\end{minipage}
\hfill
\begin{minipage}{0.31\linewidth}
	\centering
	\includegraphics[width=\linewidth]{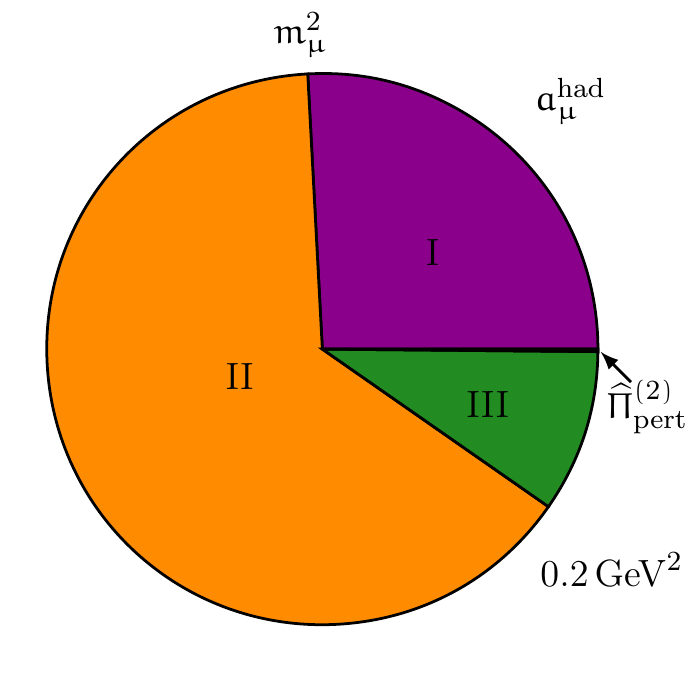}
\end{minipage}
\caption{\small{{\bf Left:} The subtracted vacuum polarization $\hat{\Pi}^{(2)}(\hat{q}^2)$
computed on the F6 ensemble. The
blue solid is the result from fit {\it ansatz} A matched to two-loop
perturbation theory (light blue line). {\bf Right:} The different contributions
to $a_\mu^\mathrm{HLO}$ broken down to different momentum ranges (see text), indicated in both panels 
by corresponding colours.}}
\label{vacpol}
\end{figure}

How statistical errors of order 5-10\% come about can be easily understood from Fig.~\ref{vacpol}.
In the left panel we show the result for $\hat{\Pi}^{(2)}(\hat{q}^2)$ together with the fit function
({\it ansatz} A) and the perturbative curve for the F6 ensemble. The horizontal axis is divided into four regions.
Region I, is defined by $0 \leq \hat{q}^2 < m_{\mu}^2$ (not visible on the left panel) 
where the $\hat{q}^2$-dependence
is only constrained by smoothness requirements on the fit function. 
There is no direct measurement of the vacuum polarization function here, although we could have tuned
the $\theta$-angle in order to penetrate this region. However, the error on the corresponding value of the
 integrand in Eq.~\ref{eq:amu}  would have turned out to be larger than 100\%.
Region II, where $m_\mu^2 \leq \hat{q}^2 < 0.2 \, {\rm GeV}^2$, is accessible only thanks to the use of  partially twisted
boundary conditions; $0.2 \, {\rm GeV}^2$ indeed coincides with the smallest Fourier mode on the F6 ensemble.
Region III,  $ 0.2 \, {\rm GeV}^2 \leq \hat{q}^2 < \hat{q}^2_{\rm max}$, is the region which is accessible
here by Fourier modes.
Region IV, $ \hat{q}^2 \geq \hat{q}^2_{\rm max}$, is the ``perturbative'' region. 
The right panel shows the relative contributions from the different regions to the integral in Eq.~\ref{eq:amu}.
Since region I contributes about 25\% and is not constrained by any direct measurement of $\Pi^{(2)}(\hat{q}^2)$,
its uncertainty dominates the overall statistical error on
$a_\mu^{\rm HLO}$, with an ambiguity which, within this approach, inevitably amounts to about
5\%. The uncertainty on the contribution from region II is of the same order but slightly smaller, whereas
the uncertainties on regions III and IV are sub-dominant.
\subsection{Chiral extrapolations}
\begin{figure}[htb]
\begin{minipage}{0.49\linewidth}
	\centering
	\includegraphics[width=\linewidth]{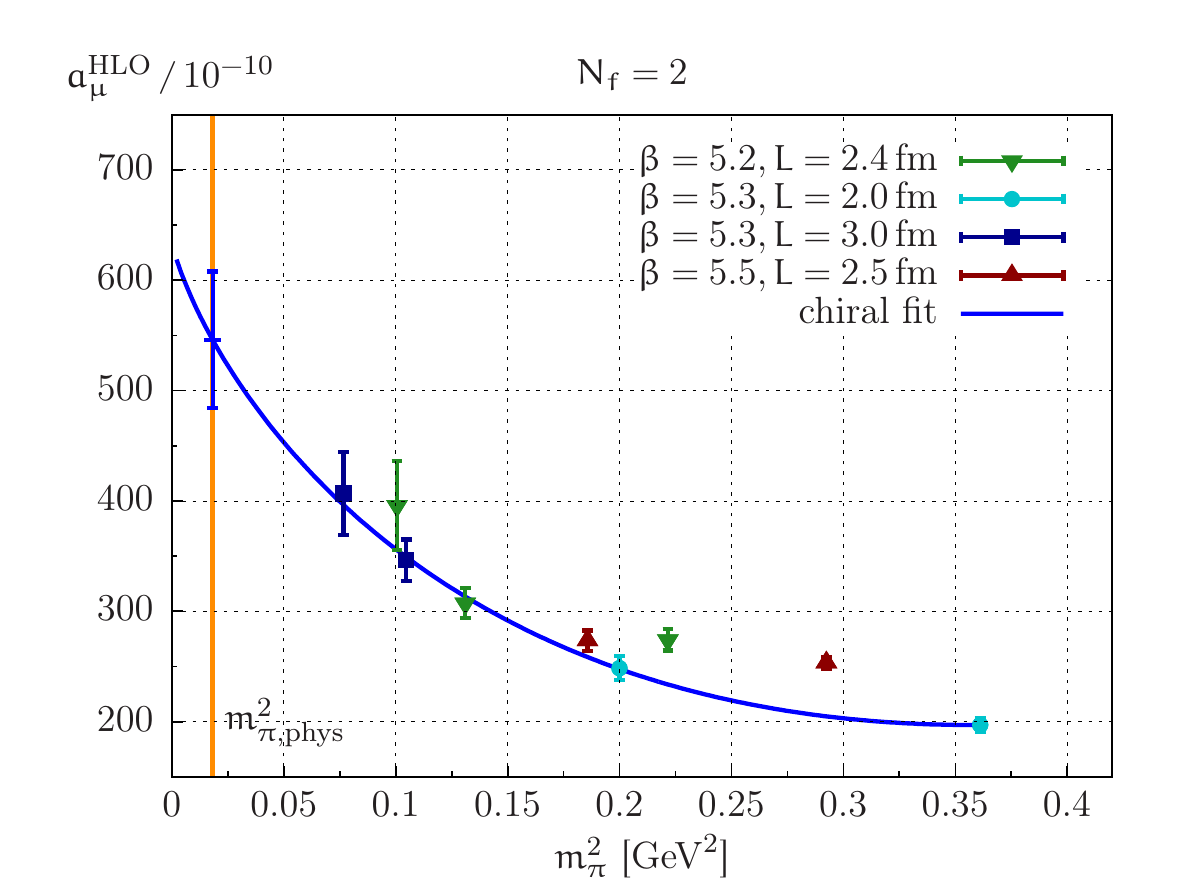}
\end{minipage}
\hfill
\begin{minipage}{0.49\linewidth}
	\centering
	\includegraphics[width=\linewidth]{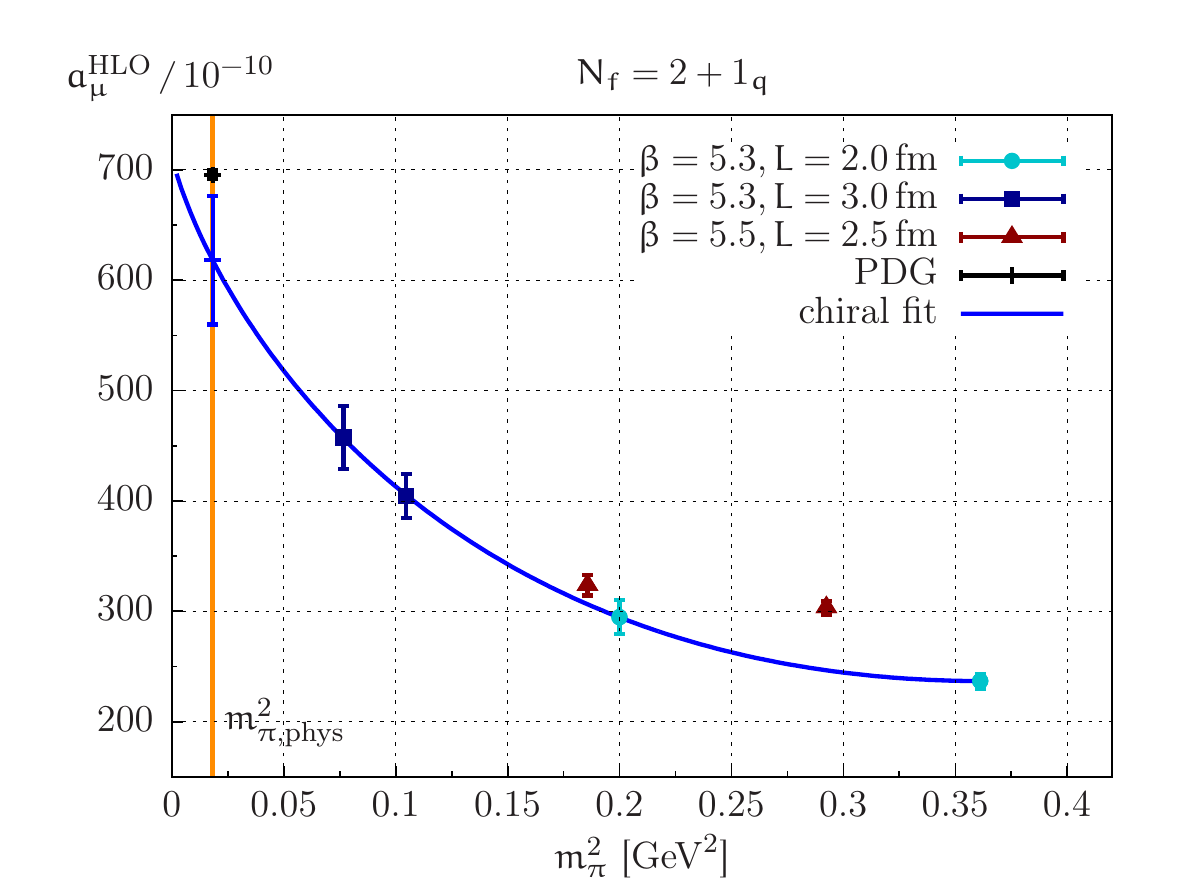}
\end{minipage}
\caption{\small{{\bf Left:} The simulation results for the hadronic contribution
to $a_\mu$ computed using two flavours, shown as a function of $m_\pi^2$. The
chiral extrapolation (blue curve) is performed for the ensembles at $\beta=5.3$
using an {\it ansatz} motivated by
chiral perturbation theory. 
{\bf Right:} Corresponding results for
$a_\mu^\mathrm{HLO}$ including a quenched strange quark. The value from the PDG~\cite{PDG} is also shown for illustration.}}
\label{chiral}
\end{figure}

We show our results for $a_\mu^{\rm HLO}$ in the $N_{\rm f}=2$ and the $N_{\rm f}=2$ plus a quenched strange quark theories
as functions of $m_\pi^2$ in Fig.~\ref{chiral}. The blue curves represent our preferred chiral fits and we will
discuss in the following how those have been obtained. We start from the observation that in the mass-region
explored a curvature is clearly visible in our data. Indeed a linear fit in that region would fail in producing
acceptable $\chi^2/$dof values. We take the observed curvature into account in three different ways.
\begin{itemize}
\item We perform a chiral fit inspired by the functional forms derived at NLO in $\chi$PT for different well
known chiral corrections. These include a chiral logarithm, which should account for the curvature.
\item We re-analyze our data following the alternative procedure presented in~\cite{Feng:2011zk}, which
is expected to remove the curvature due to the lowest lying vector resonance.
\item We impose a cut on the pion mass ($< 400$ MeV) and linearly extrapolate our data at the four lightest
values of $m_\pi$ in the $N_{\rm f}=2$ theory. 
In doing so we negelect cutoff effects and fit the $\beta=5.2$ and $\beta=5.3$ results
simultaneously. Such a procedure produces a value for $a_\mu^{\rm HLO}$ at the physical point which is well
consistent with the result from the $\chi$PT-inspired fit ${\it ansatz}$ in Eq.~\ref{Nf2res} although
slightly lower (namely $a_\mu^\mathrm{HLO}=  508\, (62)$).
\end{itemize}
We do not further discuss the third analysis and present in detail the approaches used for the first two
fitting procedures.

Since little is known about the functional form describing the dependence of $a_\mu^{\rm HLO}$ on the light quark
masses, we conservatively adopt a form inspired by $\chi$PT to extrapolate our data to the physical point.
Namely we use the fit function
\be
a_\mu^{\rm HLO}(m_\pi) = a_\mu^{\rm HLO}(0) + B\, (am_\pi)^2 + C\, (am_\pi)^2 \ln((am_\pi)^2) \;,
\ee
with $a_\mu^{\rm HLO}$, $B$ and $C$ as free parameters. Such a functional form can be derived for the
connected part of $a_\mu^{\rm HLO}$ from the expressions for the (connected) vacuum-polarization obtained
in~\cite{DellaMorte:2010aq} at NLO in $\chi$PT. 
In addition, since our current set of ensembles
 covers a wide range of pion masses
at only one value of the lattice spacing ($a=0.063$ fm), and in order to avoid mixing of cutoff effects with chiral effects,
we decide to extrapolate the $\beta=5.3$ data only. The other data sets in Table~1 are used to asses
lattice artifacts and finite size effects.
Eventually, extrapolations to the continuum limit at fixed
pion masses will have to be performed before the chiral extrapolation. 

The resulting curves are shown in Fig.~\ref{chiral}.
We find that the function describes the whole set of data
points quite well, even those which are not included in the fit. 
In the two-flavour case we obtain at the physical point
\begin{equation}
a_\mu^\mathrm{HLO}=  546\, (62) \cdot 10^{-10} \quad [N_{\rm f}=2]\;,
\label{Nf2res}
\end{equation}
which is well consistent with the recent $N_{\rm f}=2$ result in~\cite{Feng:2011zk}, 
whereas the inclusion of a quenched strange quark gives
\begin{equation}
a_\mu^\mathrm{HLO}  = 618\,(58) \cdot 10^{-10} \quad [N_{\rm f}=2+1_{\rm q}]\;,
\label{Nf2qres}
\end{equation}
where the errors are purely statistical.

In~\cite{Feng:2011zk} an alternative extrapolation procedure was proposed, with the aim of reducing 
the $m_\pi$-dependence. It can be motivated starting from a vector dominance description of $\hat{\Pi}^{(N_{\rm f})}(\hat{q}^2)$, i.e.
\be
\hat{\Pi}^{(N_{\rm f})}(\hat{q}^2,m_\pi) \propto g_{\rm V}^2 {{\hat{q}^2}\over{m_{\rm V}^2(m_\pi)+\hat{q}^2}} \;,
\label{eq:vectdom}
\ee
where the $m_\pi$-dependence is explicitly indicated. The quantity 
$g_{\rm V}$ is related to the vector decay constant $f_{\rm V}$ by 
$g_{\rm V}=f_{\rm V}/m_{\rm V}$. The dependence of $g_{\rm V}$ on $m_\pi$ is neglected based on the
numerical observation in Fig.~1 of~\cite{Renner:2010zj}. Eq.~\ref{eq:vectdom} then leads to
\be
a_\mu^{\rm HLO} \propto \int_0^\infty dq^2 f(q^2) \hat{\Pi}^{(N_{\rm f})}(q^2,m_\pi) \propto
g_{\rm V}^2 {{m_\mu^2}\over{m_{\rm V}^2(m_\pi)}}\;.
\ee
It is easy to see that, by rescaling the argument of the function $f$ from $q^2$  to $hq^2$, one obtains
\be
a_{\mu,h}^{\rm HLO} \propto g_{\rm V}^2 {{m_\mu^2}\over{hm_{\rm V}^2(m_\pi)}}\;.
\ee
Under these assumptions the choice $h=\frac{m_\rho^2}{m_{\rm V}^2(m_\pi)}$ in~\cite{Feng:2011zk} would therefore
remove the dependence on $m_\pi$ (up to the mild one in $g_{\rm V}$) producing in addition the right physical
result since $h \to 1$ in that limit. Such a rescaling has the further advantage of 
making the dimensionless quantity $a_{\mu,h}^{\rm HLO}$ independent from the lattice 
spacing~\cite{Feng:2011zk}, as opposite to $a_{\mu}^{\rm HLO}$ where the lattice spacing
value is needed to convert $m_\mu$ in the weight-function $f$ (see Eq.~\ref{eq:f}) to lattice units.

\begin{figure}[htb]
\begin{minipage}{0.49\linewidth}
	\centering
	\includegraphics[width=\linewidth]{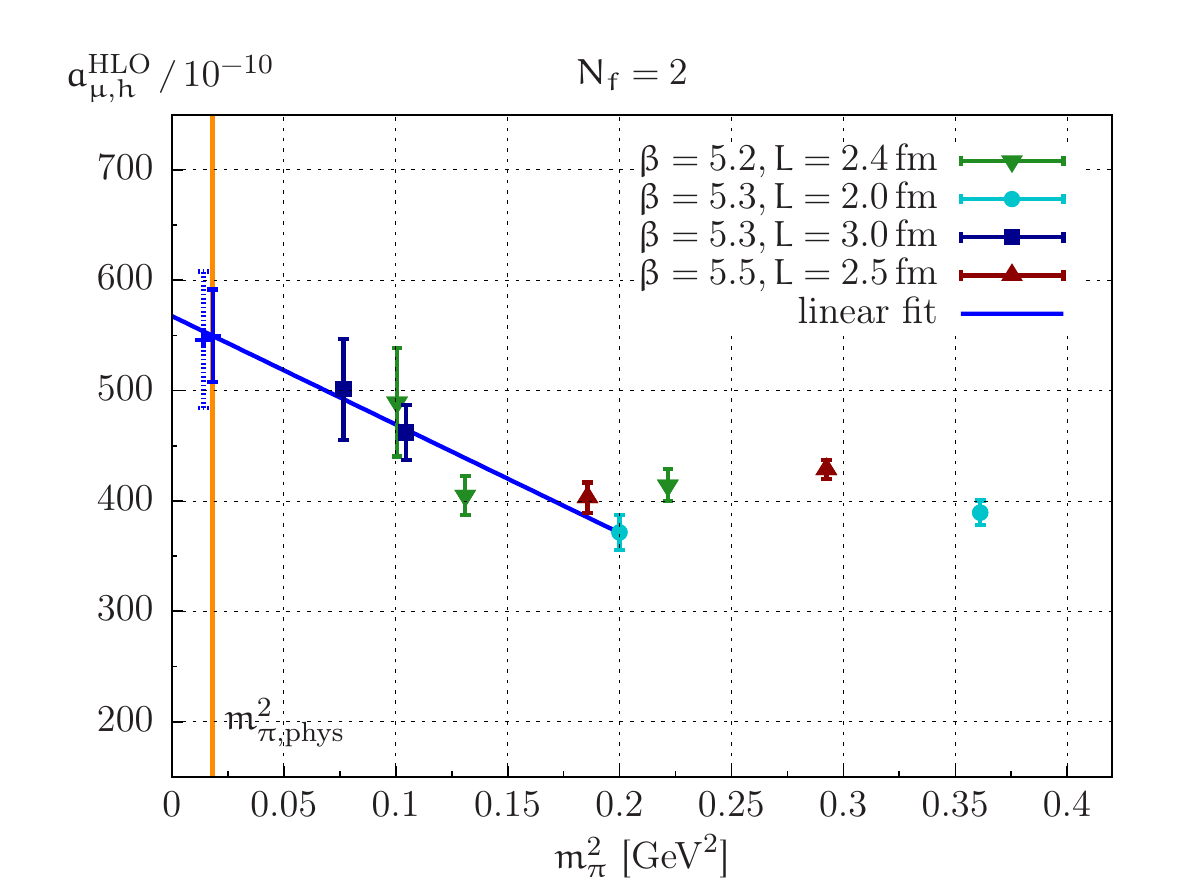}
\end{minipage}
\hfill
\begin{minipage}{0.49\linewidth}
	\centering
	\includegraphics[width=\linewidth]{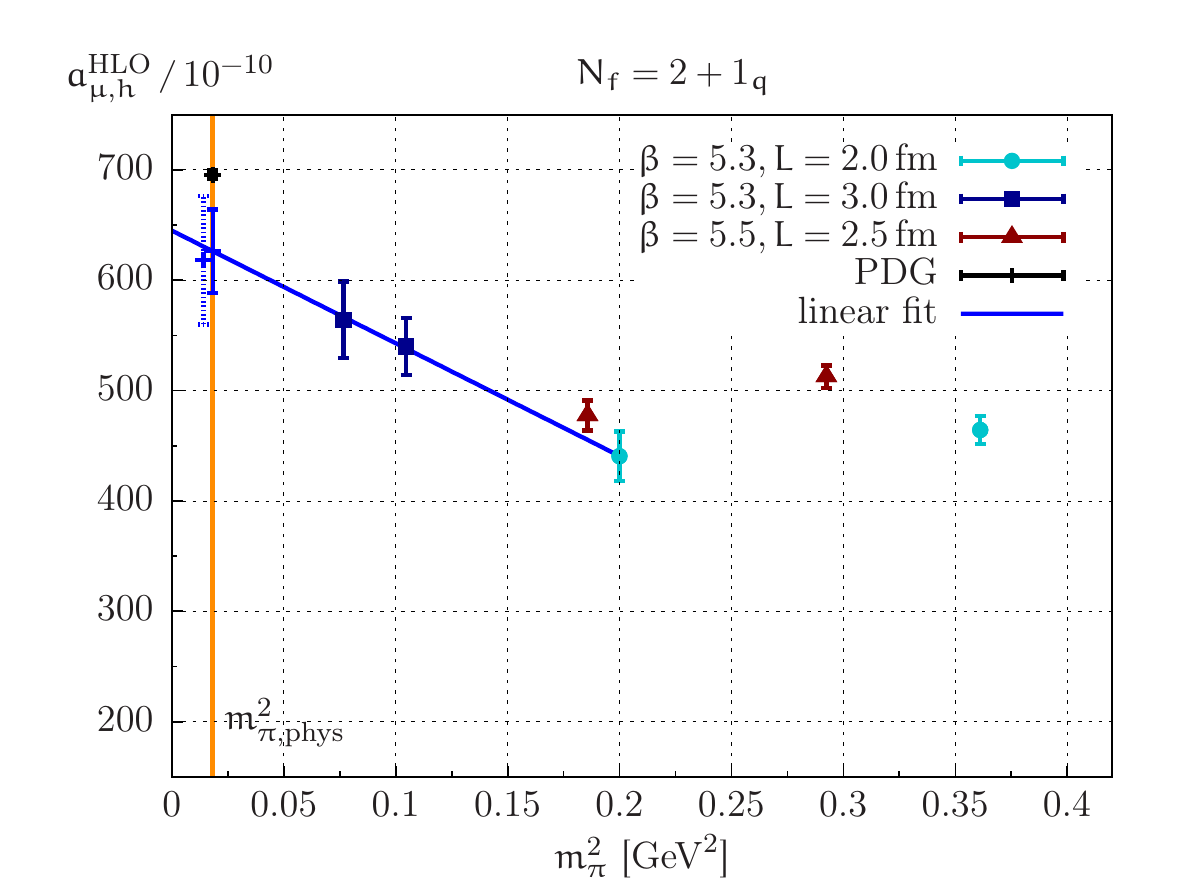}
\end{minipage}
\caption{\small{Results and chiral extrapolations for  $a_{\mu,h}^{\rm HLO}$
with $h=\frac{m_\rho^2}{m_{\rm V}^2(m_\pi)}$ 
{\bf Left:} $N_{\rm f}=2$. {\bf Right:} $N_{\rm f}=2+1_{\rm q}$.}
In both cases the leftmost value shows the result from the corresponding extrapolation in Fig.~\ref{chiral}.}
\label{chiralTM}
\end{figure}

In order to estimate the uncertainty in our chiral extrapolation we perform the analysis described above, using
the {\it naive} vector mass to define $h$. Results for $a_{\mu,h}^{\rm HLO}$ are shown in Fig.~\ref{chiralTM}.
We expect from the observation in the previous section on the high $\chi^2/$dof values produced by the fit {\it ansatz} B (single
vector dominance) that some dependence on $m_\pi$ will remain.  
However, the rescaling clearly renders such a  dependence mild, its effect being particularly strong for heavy pion masses. 
The curvature visible in Fig.~\ref{chiral} has almost completely disappeared and we therefore linearly extrapolate 
the results from the three smallest pion masses at $\beta=5.3$ to the physical point. 
There is no obvious theoretical reason why a non-linear dependence could not survive,
however, we do not have a large sensitivity to those terms in the $m_\pi$-range explored here.
We obtain in this way the following results at the physical point
\begin{equation}
a_\mu^\mathrm{HLO}  =  550\,
(42) \cdot 10^{-10} \quad [N_{\rm f}=2]\;,
\label{ETMres}
\ee
and
\be
a_\mu^\mathrm{HLO}  =  626\,
(38) \cdot 10^{-10}  \quad [N_{\rm f}=2+1_{\rm q}]\;,
\label{ETMresq}
\end{equation}
which are consistent with those in Eqs.~\ref{Nf2res}~and~\ref{Nf2qres}. Notice that, as a consequence of the
rather moderate chiral extrapolation performed, they are also compatible
with the values of $a_{\mu,h}^{\rm HLO}$ directly estimated at our most chiral point (F7 ensemble).
\subsection{Residual cutoff and finite size effects}
\label{subFSE}
The set of our simulations covers a rather wide range of lattice volumes and lattice spacings. We are therefore
able to address the issue of finite volume effects and cutoff effects. This study is not yet complete
and should be extended to lower pion masses.
\subsubsection{Cutoff effects}
\begin{figure}[htb]
\begin{center}
\includegraphics[width=8.5cm]{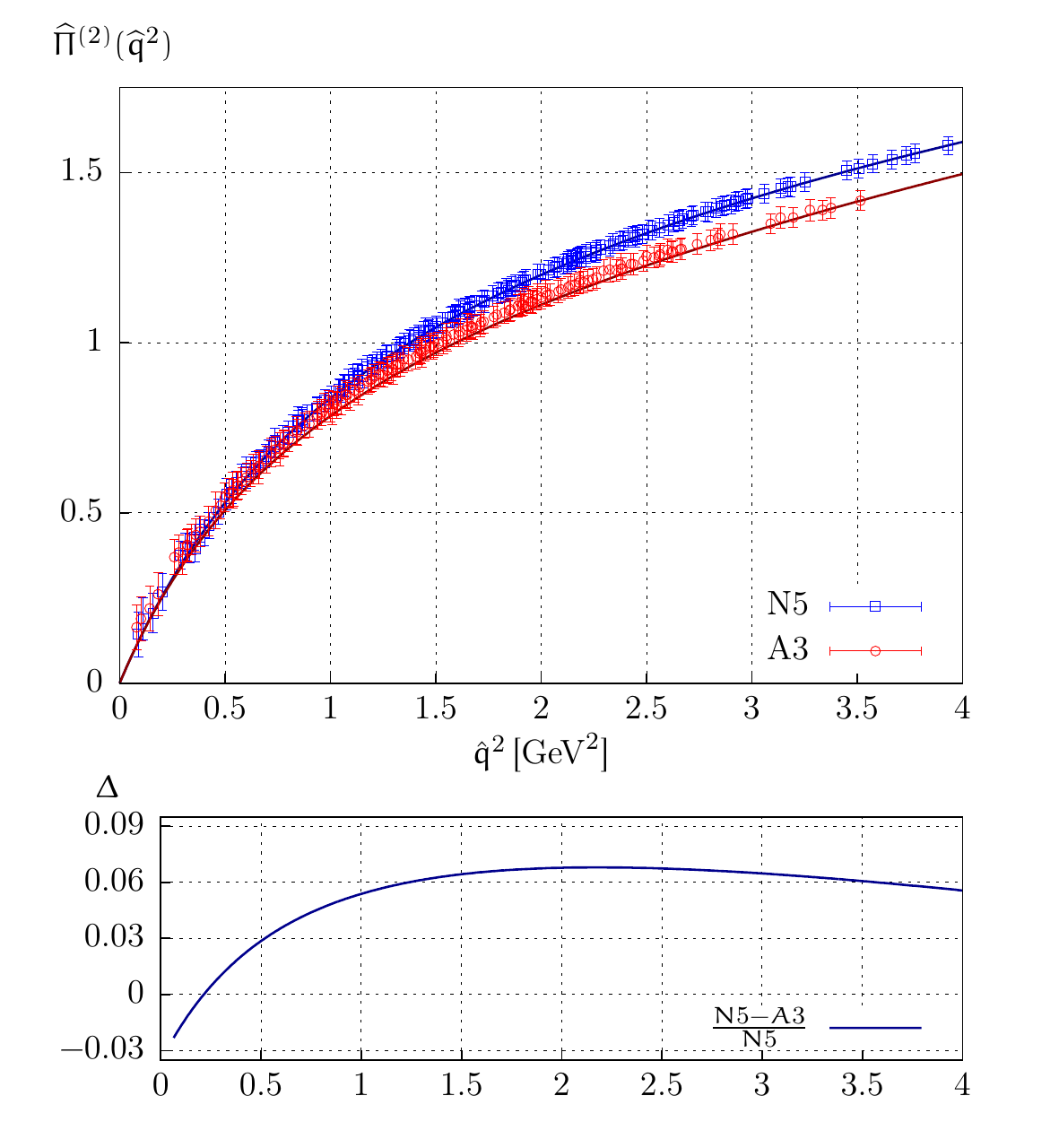}
\caption{\small{The subtracted vacuum polarization from two ensembles
with roughly the same pion mass $\m_\pi \simeq 450$ MeV and size $L \simeq 2.5$ fm, but different
lattice resolutions ($a(A3)=0.079$ fm vs. $a(N5)=0.05$ fm).}}
\label{cutoff}
\end{center}
\end{figure}

In general we expect O($a$) discretization
effects in $a_{\mu}^{\rm HLO}$
as the vacuum polarization receives off-shell contributions and the lattice 
regularization used here is only on-shell improved.
Some indications on the size of cutoff effects can be gathered from Fig.~\ref{chiral}, where at least for low pion masses
discretization effects appear to be rather small and below our statistical errors. However, 
we believe it is more instructive to look directly at lattice artifacts in
$\hat{\Pi}^{(N_{\rm f})}(\hat{q}^2)$, as those conceivably depend on $\hat{q}^2$.
In Fig.~\ref{cutoff} we compare the subtracted vacuum polarization from the ensembles N5 and A3, which have rather similar
spatial extensions and pion masses but different values of the lattice spacing.
As expected, lattice artifacts mainly distort the function at large values of $\hat{q}^2$, in the region which contributes
little to $a_\mu^{\rm HLO}$ (see Fig.~\ref{vacpol}).
\subsubsection{Finite size effects}
\begin{figure}[htb]
\begin{minipage}{0.49\linewidth}
	\centering
	\includegraphics[width=\linewidth]{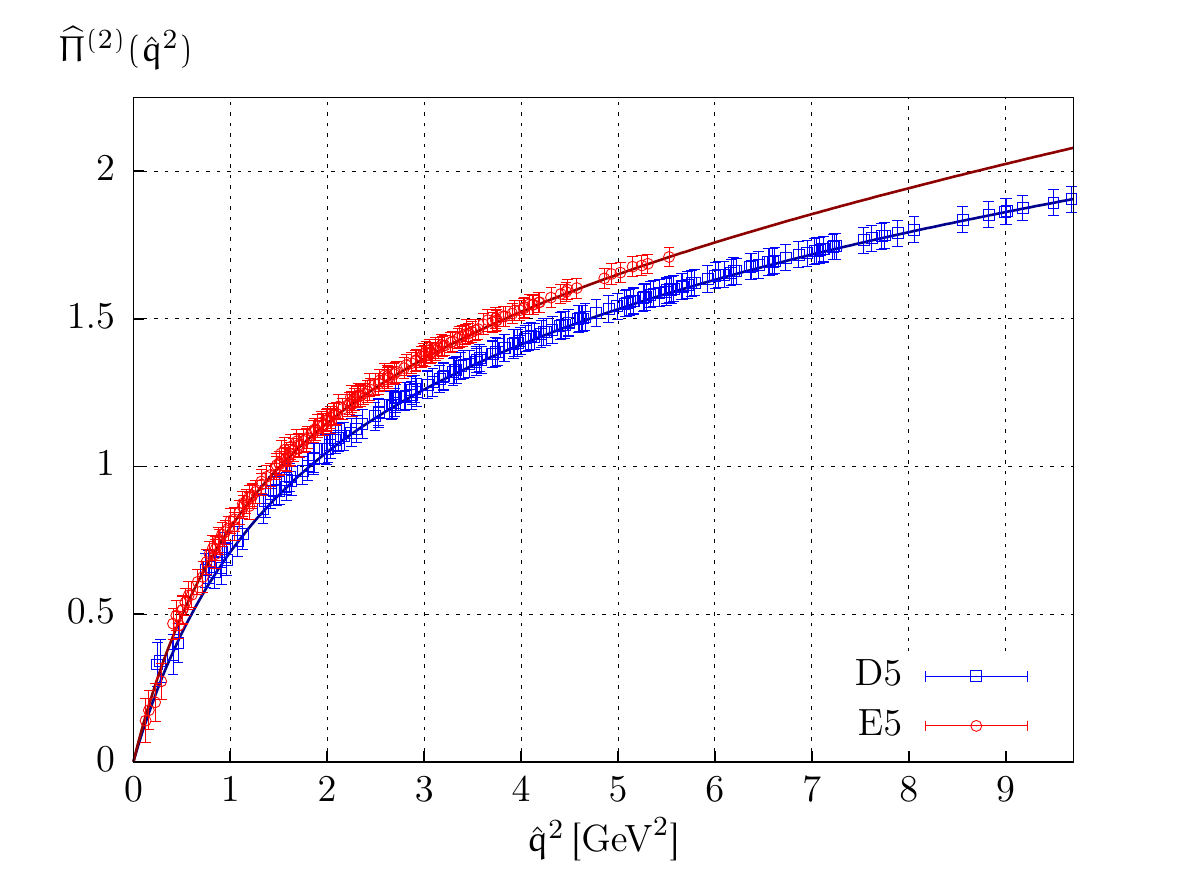}
\end{minipage}
\hfill
\begin{minipage}{0.49\linewidth}
	\centering
	\includegraphics[width=\linewidth]{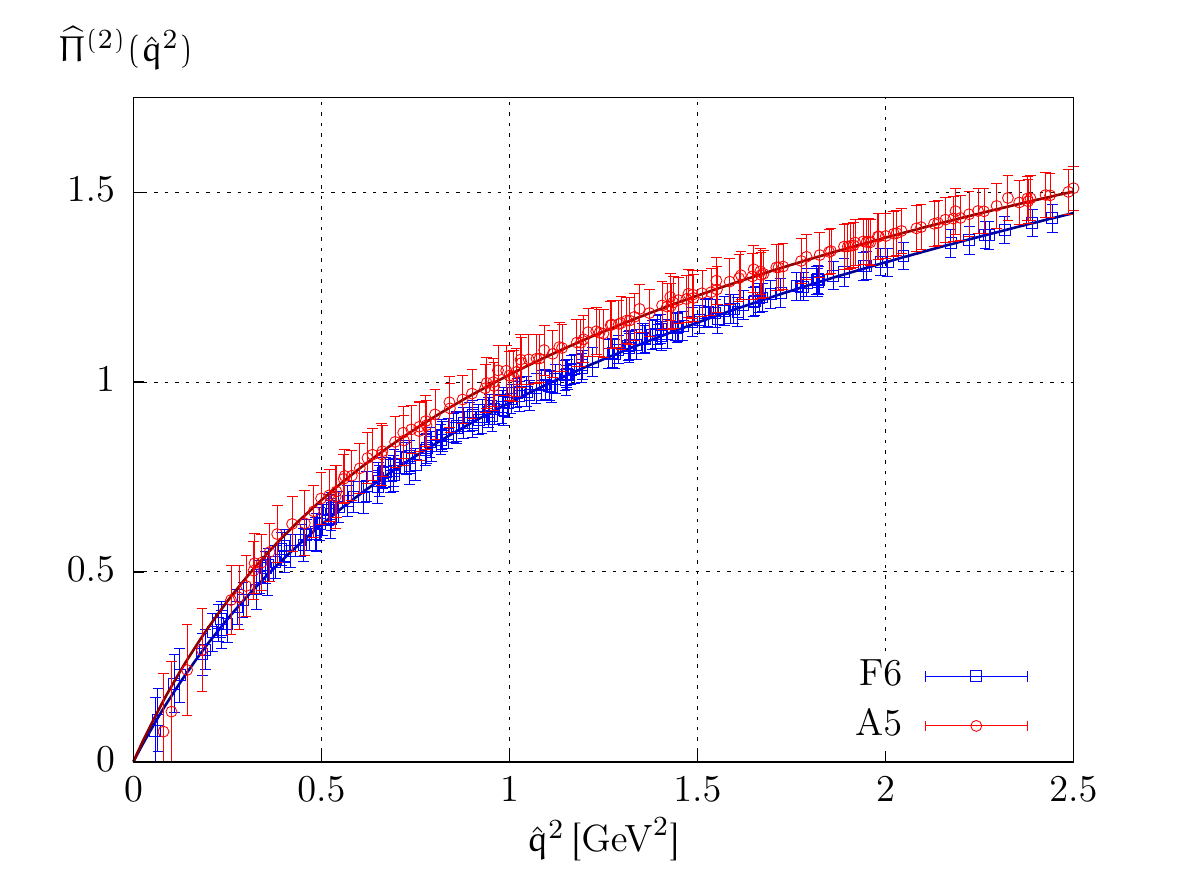}
\end{minipage}
\caption{\small{Comparison of results for $\hat{\Pi}(\hat{q}^2)$ at different values of $L$. {\bf Left:} The lattice E5 ($L/a=32$, $a=0.063$ fm)
is compared  to a $L/a=24$ lattice with the same resolution 
(lattice D5, not used here in the chiral extrapolation). {\bf Right:}
F6 ($L/a=48$, $a=0.063$ fm) vs. A5 ($L/a=32$, $a=0.079$ fm).}}
\label{FSE}
\end{figure}

As for the case of cutoff effects we look at finite size effects directly in $\hat{\Pi}^{(N_{\rm f})}(\hat{q}^2)$, since
different states are expected to contribute to the vacuum polarization when $\hat{q}^2$ is varied 
(see also~\cite{Bernecker:2011gh}). In Fig.~\ref{FSE} (left panel) we compare the results from the E5 ensemble with
those from a simulation at the same bare parameters but different $L/a$ (D5 ensemble, not in Table~1), 
specifically $L/a=24$ instead of $32$. Finite size effects are clearly visible in this rather 
extreme setup ($m_\pi L \simeq 3.4$ for D5), but they seem to drop below our present statistical error as $L$ is made larger, as the right panel of Fig.~\ref{FSE} shows. Note, however, that in 
this second comparison both $L$ and $a$ differ.

\section{CONCLUSIONS}
We have presented a calculation of the leading hadronic contribution to the anomalous magnetic
moment of the muon on the lattice with two dynamical flavours and a quenched strange quark.
We have discussed technical improvements, which led to a better determination of the external
momentum dependence of the vacuum polarization $\hat{\Pi}^{(N_{\rm f})} (\hat{q}^2)$.
Specifically, together with~\cite{Boyle:2007wg}, this paper contains on of the first numerical applications of partially twisted
boundary conditions to a quantity containing flavour-singlet currents.
The approach follows from the theoretical setup devised in~\cite{DellaMorte:2010aq}.
We restrict the computation to the connected part of  $\hat{\Pi}^{(N_{\rm f})} (\hat{q}^2)$ only.
As it is clear that presently the accuracy of lattice results is not yet at the level required 
by phenomenology, the main goals of this paper are a precise assessment of the different
sources of uncertainty and an estimate of their size. The main results are listed
in  Eqs.~\ref{Nf2res}~and~\ref{Nf2qres}, where statistical errors only are included.
In the following we estimate  the uncertainties due to the modelling
of the $q^2$-dependence of the vacuum polarization, chiral extrapolations, lattice artifacts and 
finite volume effects. These uncertainties are quantified by 
changing one ingredient at a time in the
conservative analysis procedure, which we have followed in producing our main results.

In detail, on top of the statistical error we identify the following sources:
\begin{itemize}
\item {\bf Fitting procedure.} We repeat the entire analysis by using fit {\it ansatz} C instead of {\it ansatz} A 
everywhere and obtain
\begin{equation}
a_\mu^\mathrm{HLO} =  549\,
(55) \cdot 10^{-10}\quad [N_{\rm f}=2]\;, \quad {[{\it ansatz}\; {\rm C}]} \;,
\ee
\be
 a_\mu^\mathrm{HLO} =  615\,
(56) \cdot 10^{-10}\quad [N_{\rm f}=2+1_{\rm q}]\;, \quad {[{\it ansatz}\; {\rm C}]}\;,
\end{equation}
which suggests an uncertainty well below our statistical errors.
\item {\bf Chiral extrapolation.} By comparing the results in Eqs.~\ref{Nf2res}~and~\ref{Nf2qres}
to those in Eqs.~\ref{ETMres} and~\ref{ETMresq}, we conclude that at the moment this systematic 
cannot be resolved with our statistical errors. 
\item {\bf Cutoff and finite size effects.} As discussed in Sect.~\ref{subFSE} these effects 
appear to be small at the volumes, masses and  
lattice spacings considered here, but a more comprehensive study is required.
\item{\bf Uncertainty in the lattice spacing.} In~\cite{Capitani:2011fg} the lattice spacing at $\beta=5.3$
is given as $a=0.063(3)$ fm, by combining statistical and systematic errors in quadrature. By repeating
the analysis of our ($\beta=5.3$) data using $a=0.066$ fm  instead of $a=0.063$ fm we arrive at
\begin{equation}
a_\mu^\mathrm{HLO}  =  594\,
(66) \cdot 10^{-10}\quad [N_{\rm f}=2]\;, \quad {[a=0.066\; \rm fm]} \;,
\ee
\be
 a_\mu^\mathrm{HLO}  =  671\,
(64) \cdot 10^{-10}\quad [N_{\rm f}=2+1_{\rm q}]\;, \quad {[a=0.066\; \rm fm]}\;.
\end{equation}
We will include this systematic by taking half the difference with respect to the results in 
Eqs.~\ref{Nf2res}~and~\ref{Nf2qres}.
\end{itemize}
In general we see that at the moment most of these systematic effects,
with the exception of the uncertainty on $a$, are well below our statistical errors.
They will become relevant and will have to be more precisely estimated 
once the latter are reduced. 

We  quote as final results the values determined at $\beta=5.3$, having used
the other two lattice spacings to estimate systematic uncertainties. No
continuum extrapolation has been performed yet.  
The present computation of the connected contribution
to the anomalous magnetic moment of the muon then gives
\be
a_\mu^\mathrm{HLO}  =  546\,
(66) \cdot 10^{-10} \quad [N_{\rm f}=2]\;,
\ee
and
\be
 a_\mu^\mathrm{HLO}  =  618\,
(64) \cdot 10^{-10} \quad [N_{\rm f}=2+1_{\rm q}] \;,
\ee
where we have combined in quadrature statistical and systematic errors.

The overall error can definitely be reduced, as it is still statistics-dominated. 
However, once it is evaluated, the contribution from disconnected diagrams, which is estimated
to be around 10\% (see~\cite{DellaMorte:2010aq}), will become the main uncertainty.
We therefore consider studying the accuracy that can be reached on the numerical estimates of the disconnected 
contribution as a priority. A combination of the methods discussed 
in~\cite{Thron:1997iy,Foley:2005ac,Collins:2007mh,Ehmann:2009ki} 
seems very promising in this respect, and we have also started implementing similar techniques
in the context of mesonic three-point functions~\cite{Vera}.
Further improvements include considering lighter pion masses and enlarging the set of
simulations at $\beta=5.5$.

\vspace{+2mm}
\noindent{\bf{Acknowledgments:}} We thank Fred Jegerlehner, Rainer 
Sommer, Gilberto Colangelo, Achim Denig, and Harvey 
B. Meyer for 
useful discussions. We are grateful to our colleagues within the CLS
project for sharing gauge ensembles. Calculations of correlation 
functions were performed on the dedicated QCD platform "Wilson" at the
Institute for Nuclear Physics, University of Mainz.
We thank Dalibor Djukanovic for technical support.
This work was supported by DFG (SFB443) and the Research Center EMG 
funded by {\it Forschungsinitiative Rheinland-Pfalz}.

\vspace{.6cm}

\bibliographystyle{JHEP}
\bibliography{spires}

\end{document}